\documentclass[aps]{revtex4}
\usepackage{graphicx,epsfig}
\begin{document}
\title{Readdressing the hierarchy problem in a Randall-Sundrum scenario with bulk Kalb-Ramond background}
\author{Saurya Das\footnote{E-mail: saurya.das@uleth.ca}$^a$, Anindya Dey\footnote{E-mail: anindya@mri.ernet.in}$^b$, 
Soumitra SenGupta\footnote{E-mail: tpssg@iacs.res.in}$^c$}
\affiliation{$^a$Department of Physics\\
University of Lethbridge \\
4401 University Drive, Lethbridge\\
Alberta - T1K 3M4, Canada}
\affiliation{$^b$Department of Physics\\ 
Harish-Chandra Research Institute\\
Chhatnag Road, Jhusi, Allahabad - 211019, India}
\affiliation{$^c$Department of Theoretical Physics\\
Indian Association for the Cultivation of Science\\
Calcutta - 700032, India}
\begin{abstract}
We re-examine the fine tuning problem 
of the Higgs mass, when an antisymmetric two form Kalb-Ramond (KR) 
field is present in the bulk of a Randall-Sundrum (RS) 
braneworld. Taking into account the back-reaction of the KR field, 
we obtain the exact correction to the RS metric. 
The modified metric also warps the Higgs mass from Planck scale
(in higher dimension) to TeV scale (on the visible brane) for a 
range of values of $kr$ exceeding the original RS value
(where $k=$ Planck mass and $r=$ size of extra dimension). 
However, it requires an extraordinary suppression of the KR field density,
indicating the re-appearence of the fine tuning problem in a different guise. 
The new spacetime also generates a small negative cosmological constant 
on the visible brane. 
These results are particularly relevant for certain string based
models, where the KR field is unavoidably present in the bulk.  
We further show that such a bulk antisymmetric KR field 
fails to stabilize the braneworld.  

\end{abstract}
\maketitle

Despite the success of the standard model of elementary particles in explaining
physical phenomena up to the electroweak scale, 
the radiative instability of the Higgs mass remains an unresolved issue. 
The minimal supersymmetric standard model (MSSM) 
can solve this problem, but not without the introduction of a host of 
superpartners in the theory. As viable alternatives to supersymmetry, 
two models have recently been proposed, both of which 
require the presence of one or more extra spatial dimension(s). 
These models are respectively known as the ADD (named after Arkani-Hamed, 
Dimopoulos and Dvali) \cite{add} and RS 
(named after Randall and Sundrum) \cite{rs}.  
For other non-supersymmetric 
compactifications which reproduce the standard model at low
energies, see e.g. \cite{koko}.

In particular, in the RS approach, one considers 
a $5$ dimensional anti de-sitter spacetime with the extra spatial dimension 
orbifolded as $S_1/Z_2$. Two $(3 + 1)$- dimensional branes, known as the  
visible (TeV) brane and hidden (Planck) 
brane are placed at the two orbifold fixed points with the following 
bulk metric ansatz:
\begin{equation}
ds^2=\exp{(-A)}\eta_{\mu\nu}dx^{\mu}dx^{\nu} - r^2d\phi^2~,
\label{eq1}
\end{equation}
where $\mu,\nu=0,1,2,3$ (i.e. the visible coordinates) and $\phi$ is the 
hidden coordinate
($\eta_{\mu\nu}$ is the usual $4$-dimensional Minkowski metric, whereas
$G_{MN}$ etc will denote the full five-dimensional metric). 
As a consequence of the 
warped geometry, all mass scales get exponentially lowered from the 
Planck scale on the hidden brane to the TeV scale on the visible brane.
The radius of the compact dimension, on the other hand, remains close 
to the Planck length, the $4$ and $5$ dimensional Planck scales remain
close to each other and no scale hierarchy problem
appears in this case. In both the above models,
the standard model fields are assumed to be localized on the 
visible brane, whereas gravity alone propagates in the bulk. 
This is consistent with string theory, where 
the standard model fields appear as open string modes
attached to the visible brane,
whereas gravity being a closed string excitation, propagates in the bulk.
Attempts to stabilize such a braneworld include those which
incorporate a scalar field in the bulk, 
resulting in an effective potential for the
compactification radius for the RS model \cite{gw}.
Such a model however does not take into account the back-reaction 
of the scalar field on the
background metric. Several other pieces of work followed, which  
compute the scalar back-reaction on the metric \cite{backreaction}. 
But except for some special cases, an exact back-reacted solution has not been
derived.

Our present work is motivated from a different angle.
In the context of string theory,
apart from the graviton and scalar excitations, the second rank
anti-symmetric tensor field (the KR field) 
in the Neveu-Schwarz-Neveu-Schwarz (NS-NS) sector of the underlying string
theory, can also be present in the bulk as a closed string mode. Various 
aspects of the presence of the antisymmetric fields have
been discussed in the context of D-brane models \cite{burgess}. 
Moreover ref.\cite{ssg} explored various cosmological implications of the KR
field in RS background. In this paper we explore the effect of the KR
field on the background metric and re-examine the 
fine-tuning problem. 
We first derive an exact solution for the metric and exhibit 
its deviation from the RS solution. 
The new metric depends on  the energy density of
the KR field, and goes to the RS solution smoothly in the limit of 
KR energy density tending to zero. This scenario solves the 
hierarchy problem as well, not just for the value of $kr$ predicted
by RS, but for any value greater than the RS value. 
However, it requires the KR energy density to be exceedingly small, 
signalling the return of the fine tuning problem in a different guise. 
This value nonetheless
matches remarkably well with the 
KR energy density on the visible brane, calculated from 
the solution of the KR field in a 
Randall Sundrum braneworld in a previous work \cite{ssg1}.  
Moreover, the KR density induces a negative cosmological constant in the 
visible brane. 

We begin with the RS metric ansatz (\ref{eq1}), and the action 
\begin{eqnarray}
S &=& S_{Gravity} + S_{vis} + S_{hid} + S_{KR}~, \\
\mbox{where,}~~~~S_{Gravity} &=& \int d^4x~r~d{\phi} \sqrt{G}~[ 2M^3R + \Lambda]\\
S_{vis} &=&  \int d^4x \sqrt{-g_{vis}}~[L_{vis} - V_{vis}]\\
S_{hid} &=&  \int d^4x \sqrt{-g_{hid}}~[L_{hid} - V_{hid}]\\
S_{KR} &=& \int d^4x~r~d{\phi}\sqrt{G}~[ 2H_{MNL} H^{MNL}]~.
\end{eqnarray}
Here $\Lambda$ is the five dimensional cosmological constant, 
$V_{vis}, V_{hid}$ are the visible and hidden brane
tensions. 
$H_{MNL} = \partial_{[M} B_{NL]}$ 
is the third rank antisymmetric field strength 
corresponding to the two-form
KR field $B_{MN}$ \cite{ssg1}.
The $5$ dimensional Einstein equations are as follows (where $'\equiv d/d\phi$):
\begin{eqnarray}
\frac{3}{2}{A'}^2&=&-\frac{\Lambda}{4M^3}~r^2 -
\frac{3}{2M^3}
g^{\nu\beta}g^{\lambda\gamma}H_{\phi\nu\lambda}H_{\phi\beta\gamma}
\label{eq2} \\
\frac{3}{2}({A'}^2-{A''})&=&-\frac{\Lambda}{4M^3}r^2+\frac{\exp(-2A)}{2M^3}
~\eta^{\lambda\gamma}
[- 12\eta^{00} H_{\phi 0 \lambda}H_{\phi 0 \gamma}+3\eta^{\nu\beta} 
H_{\phi\nu\lambda}H_{\phi\beta\gamma}]
\label{eqna2}
\\
\frac{3}{2}({A'}^2-{A''})&=&-\frac{\Lambda}{4M^3}r^2+\frac{\exp(-2A)}{2M^3}
~\eta^{\lambda\gamma}
[-12\eta^{ii} H_{\phi i \lambda}H_{\phi i \gamma}+3\eta^{\nu\beta}
H_{\phi\nu\lambda}H_{\phi\beta\gamma}]
\label{eqna3}
\end{eqnarray}
In Eq.(\ref{eqna3}), 
the index $i$ on the right hand side runs over 1,2 and 3, i.e. three spatial 
components $x,y,z$, and there is no sum over $i$. Also, $\eta^{ij} \equiv g^{im} g^{jn} \eta_{mn}$. 
Adding Eq.(\ref{eqna2}) and the $x,y,z$ components of Eq.(\ref{eqna3}),we get,
\begin{equation}
{A'}^2-{A''}=-\frac{\Lambda}{6M^3}r^2
\label{eineqn6} 
\end{equation}
\noindent
which has the solution ($k \equiv \sqrt{-\Lambda/24M^3}~;~b,c=$ integration constants): 
\begin{equation}
\exp(-A)=\frac{\sqrt{b}}{2kr}\cosh{(2kr\phi+2krc)}~,
\label{warp2}
\end{equation}
Although $H_{\mu\nu\lambda}$ terms seem to have disappeared from
Eq.(\ref{eineqn6}), it follows from Eq.(\ref{eq2}) that 
the parameter $b$ is proportional to the energy density of the 
KR field. The constant $c$ needs to be determined from the boundary conditions.
Imposing the condition $A(\phi=0)=0$ in Eq.(\ref{warp2}), we get:
\begin{equation}
\frac{2kr}{\sqrt{b}}=\cosh(2krc)~.
\label{sinh1}
\end{equation}
In the limit $c \rightarrow -\infty$, Eq.(\ref{warp2}) becomes:
$$\exp(-A) = \frac{\sqrt{b}}{4kr} \exp(-2krc) \exp(-2kr\phi) ~,$$
while Eq.(\ref{sinh1}) becomes:
$$ \frac{4kr}{\sqrt{b}} = \exp(-2krc) ~.$$
(This shows that $b\rightarrow 0$, i.e. the KR field vanishes 
in the above limit).
Substituting the latter in the former, we get:
$$ \exp(-A) = \exp(-2kr\phi)~,$$
which is the Randall-Sundrum result for the warp factor. 
%
%
%
%
%
%
Further, using Eq.(\ref{warp2}) and the following condition near $\phi=0,\pi$:
\begin{eqnarray}
{A''} &=& 
\frac{r}{6M^3}
\left[
~V_{hid}~\delta(\phi) + 
{V_{vis}}~\delta(\phi-\pi)
~\right]~,
\\
\mbox{we get},~~~ 
c &=& - \frac{1}{2kr} \tanh^{-1} \left( \frac{V_{hid} }{24 M^3 k} \right)
=  -\pi + \frac{1}{2kr} \tanh^{-1} \left( \frac{V_{vis} }{24 M^3 k} \right)~,
\end{eqnarray}
or equivalently:
\begin{equation}
V_{vis} 
=
24 M^3 k \tanh\left( 2kr (\pi + c) \right)
=-\frac{\Lambda}{k} \tanh\left( 2kr (\pi + c) \right)
~,~~
 V_{hid} = 
-24 M^3 k \tanh\left( 2krc \right) 
=\frac{\Lambda}{k} \tanh\left( 2krc \right) ~.
\label{vvis1}
\end{equation}
Once again, note that in the $c\rightarrow - \infty$ limit, 
one gets $V_{hid}=-V_{vis}=24M^3k$, as in \cite{rs}.
Once the metric is determined, we 
examine the condition under which the desired reduction of the Higgs mass 
from the Planck scale to the TeV scale can be achieved. 
From
the warp factor given by Eq.(\ref{warp2}) and 
Eq.(\ref{sinh1}), it follows that the Higgs mass at the visible brane is:
\begin{equation}
m_{H}^2=\frac{\sqrt{b}}{2kr}\cosh\left[2kr\pi + \cosh^{-1} \frac{2kr}{\sqrt{b}}\right]m_{0}^2
=
\left[
\cosh\left(2 kr\pi\right) - \sinh\left( 2kr\pi\right)
\sqrt{1- \frac{b}{(2kr)^2}}
~\right]
m_0^2~,
\label{mh1}
\end{equation}
where $m_{H} \approx$ TeV and $m_{0}\approx 10^{16}$ TeV
denotes the mass parameter on the Planck (hidden) brane, such that 
$m_H/m_0=10^{-16}$.
As expected, in the limit $b\rightarrow 0$, the relation $m_H = \exp(-kr\pi)~m_0$ 
is recovered, from which one obtains
$kr = (16/\pi)\ln(10)=11.7269\dots $. We will call this the RS value of $kr$.
Eq.(\ref{mh1}) can be inverted to obtain:
\begin{equation}
b = (2kr)^2
\left[1 -
\left(  
\coth(2kr \pi) - (m_H/m_0)^2 {\rm cosech}(2kr\pi)
\right)^2 
\right]~.
\label{bsolution}
\end{equation}
It follows from the last equation that $b=0$ at the 
RS value of $kr$, $b<0$ for $kr<$ the RS value and 
$b>0$ for $kr>$ the RS value. 
For $kr \gg $ RS value, $b \rightarrow 0$, 
whereas for $kr \rightarrow 0$, the asymptotic value
of $b =-1/\pi^2 = -0.1013\dots$ is reached. 
Since positivity of 
$b$ is required for the metric to be real (see Eq.(\ref{warp2})), 
we conclude that for a non-vanishing KR field, however small, 
$kr$ has to exceed the RS value. 
The plot of $\log|b|$ vs. $kr$ is shown in Fig.1. 
Note that there is no upper bound on $kr$, 
and in principle it can be as large as desired. 
In other words, the hierarchy problem can be solved for 
any value of $kr>$ the RS value, the corresponding value of the 
KR field given by Eq.(\ref{bsolution}). 
The kink in the graph corresponds to the RS value, when $b=0$ and it can be seen that 
as $kr$ increases from the RS value, $b$ reaches a maximum of about $10^{-61}$
and then falls rapidly to zero. 
In other words, 
the KR field in the observable universe has to be exceedingly small!
It is 
interesting to note that $b \approx 10^{-62}$ was also
predicted in \cite{ssg1}. There, 
the KR field in the bulk was decomposed 
into various Kaluza-Klein (KK) modes and the equations of motion for these 
modes were obtained. The solutions and the masses for each of these KK modes 
then followed from these equations, and one
obtained $H_{\mu\nu\lambda} \approx M^{3/2} 10^{-31}$, from which it followed that
$b \approx 10^{-62}$, similar to what we obtained here!

\begin{figure}[t]
\begin{center}
\epsfxsize 3.40 in
\epsfbox{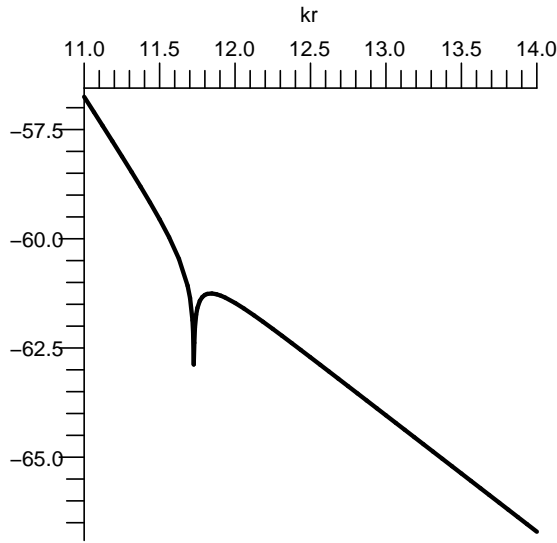}
\caption{Plot of $\log|b|$ vs. $kr$ for $kr=11 \dots 14$. 
The kink corresponds to $kr=11.7269\dots$, 
the RS value, where
$b=0$ and $\log|b| \rightarrow -\infty$. Above the RS value, $b$ rises to a maximum of 
$10^{-62}$ and then decays to zero. 
Below the RS value, $b < 0$, and the corresponding metric
is unphysical.
The curve meets the $\log|b|$ axis at $\log(1/\pi^2) = -0.9943$ (not shown). 
}
\end{center}
\end{figure}

Finally, we compute the effective four dimensional cosmological constant 
on the visible brane \cite{maar}. From (\ref{vvis1}), we get:
\begin{eqnarray}
\lambda \equiv \frac{1}{2}
\left( k V_{vis} + \Lambda \right)
&=&  12 M^3 k \left[
\tanh(2krc ) + 1  \right] \\
&\approx& 
-24 M^3 k~\exp(4krc) ~
\label{lamb1}
\\
&=& 
-24 M^3 k~\frac{b}{(4kr)^2} ~
\label{lamb2}\\
&=& 
-6 M^3 k 
\left[1 -
\left(  
\coth(2kr \pi) - (m_H/m_0)^2 {\rm cosech}(2kr\pi)
\right)^2 
\right]~.
\label{lamb3}
\end{eqnarray}
where the limit $c\rightarrow -\infty$ has been taken to obtain Eq.(\ref{lamb1}),
Eq.(\ref{sinh1}) has been used in the same limit to obtain Eq.(\ref{lamb2}) and 
Eq.(\ref{bsolution}) has been used to obtain (\ref{lamb3}). 
Note that the cosmological constant gets determined in terms on $kr$ alone.
However, for any non-zero $b>0$, or $kr > $ RS value, its signature is
negative and it attains a minimum value of about 
$-10^{-63}$ (in Planck units). 
Thus, if we take the accepted value 
of $\lambda \approx 10^{-123}$ (in Planck units) seriously, 
it appears that one would have to invoke other fields to 
cancel the negative contribution to $\lambda$ from the KR field.

%

Finally, to explore whether the KR field 
helps to stabilize the braneworld, we follow the prescription of  
\cite{gw}. 
Remarkably we do not need the solution for the KR field to obtain the induced 
potential that may help in stabilization. 
We take Eq.(\ref{eq2}) along with the solution 
for the metric, plug it back in the
KR action, and integrate over the compact coordinate $\phi$, 
to arrive at the following effective potential, 
\begin{equation}
V=-12\pi \frac{M^{3}b}{r}~.
\end{equation}
Clearly, this potential does not have a minimum 
for $r\neq 0$, indicating that 
stabilization of the braneworld by KR fields may not be possible. 

We conclude with the following observations.
Presence of a bulk KR field yields a new warped solution for the metric which 
is also capable of suppressing the Higgs mass on the visible brane,
provided 
(i) $kr$ exceeds the value originally predicted by RS but otherwise 
can remain unbounded, and
(ii) the parameter $b$, representing the KR field, is heavily suppressed. 
We interpret the second requirement as the effective 
re-appearence of the fine tuning problem. 
Furthermore, the KR field generates an effective negative
cosmological constant on the visible brane. 
Stabilization of the braneworld however, cannot be achieved even if 
the KR field is considered in the bulk along with gravity.  
We hope to further 
report on these and related issues elsewhere \cite{ddsg2}.

\vspace{.1cm}
\noindent
{\bf Acknowledgement}\\
SD thanks A. Dasgupta for discussions. SD and AD thank 
the Indian Association for the Cultivation of Science for hospitality, where 
part of the work was done. 
This work was supported by the Natural 
Sciences and Engineering Research Council of Canada and the 
University of Lethbridge.


\begin{thebibliography}{99}

\bibitem{add}
N. Arkani-Hamed, S. Dimopoulos, G. R. Dvali,
Phys. Lett. {\bf B429} (1998) 263 (hep-ph/9803315);
I. Antoniadis, N. Arkani-Hamed, S. Dimopoulos, G. R. Dvali,
Phys. Lett. {\bf B436} (1998) 257 (hep-ph/9804398).


\bibitem{rs}
L. Randall, R. Sundrum, 
Phys. Rev. Lett. {\bf 83} (1999) 3370 (hep-ph/9905221);
ibid., 4690 (1999) (hep-ph/9906064). 

\bibitem{arkani1} N. Arkani-Hamed, S. Dimopoulos, 
J. March-Russell, Phys. Rev. {\bf D63} (2001) 064020 (hep-th/9809124).

\bibitem{koko} C. Kokorelis, 
Nucl.Phys. B677 (2004) 115-163 (hep-th/0207234). 

\bibitem{gw} 
W. D. Goldberger, M. B. Wise, 
Phys. Rev. Lett. {\bf 83} (1999) 4922 (hep-ph/9907447). 
J.Garriga, O.Pujolas and T.Tanaka, 
Nucl. Phys. {\bf B605} (2001) 192 (hep-th/0004109).

\bibitem{backreaction} 
C. Csaki, M. Graesser, L. Randall and J. Terning, Phys. Rev. {\bf D62} (2000) 045015 
(hep-ph/9911406);
C. Csaki, M. Graesser and G. Kribs, 
Phys. Rev. {\bf D63} (2001) 065002 (hep-th/0008151);
O. DeWolfe, D. Freedman, S.S. Gubser and A. Karch, Phys. Rev. {\bf D62} (2000) 046008 
(hep-th/9909134).

\bibitem{burgess} C. P. Burgess, C. Nunez, F. Quevedo, G. Tasinato, I. Zavala,
JHEP 0308:056 (2003) (hep-th/0305211).

\bibitem{ssg} D. Maity, S. SenGupta, S. Sur, Phys. Rev. {\bf D72} (2005) 066012
(hep-th/0507210);
B. Mukhopadhyaya, S. Sen, S. Sen, S. SenGupta Phys. Rev. {\bf D70} (2004) 066009
(hep-th/0403098); 
D. Maity, P. Majumdar S. SenGupta, JCAP 0406:005 (2004)
(hep-th/0401218);
D. Maity, S. SenGupta, Class. Quant. Grav. {\bf 21} 3379 (2004) (hep-th/0311142).

\bibitem{ssg1} B. Mukhopadhyaya, S. Sen, S. SenGupta,  
Phys. Rev. Lett. {\bf 89} (2002) 121101; Erratum-ibid. {\bf 89} (2002) 259902
(hep-th/0204242). 

\bibitem{maar} R. Maartens,
``Brane-World Gravity'',
{\it Living Rev. Relativity}, 
{\bf 7}, (2004), 7.
[Online Article]: cited [10 Feb 2006], 
http://www.livingreviews.org/lrr-2004-7 . 

\bibitem{ddsg2} S. Das, A. Dey, S. SenGupta (work in progress). 

\end{thebibliography}
\end{document}